\documentclass[12pt]{article}
\usepackage{array,multicol}
\usepackage{afterpage,float,flafter}
\usepackage[dvips]{epsfig,rotating}
\usepackage{pifont,fancybox}

\newcommand{\be}{\begin{equation}}
\newcommand{\ee}{\end{equation}}

\newcommand{\bea}{\begin{eqnarray}}
\newcommand{\eea}{\end{eqnarray}}
\newcommand{\eq}[1]{eq.~(\ref{#1})}

\newcommand{\gsim}{\ \rlap{\raise 2pt\hbox{$>$}}{\lower 2pt \hbox{$\sim$}}\ }
\newcommand{\lsim}{\ \rlap{\raise 2pt\hbox{$<$}}{\lower 2pt \hbox{$\sim$}}\ }

\newcommand{\matr}{\left( \begin{array}}
\newcommand{\ematr}{\end{array} \right)}

\newcommand{\np}[1]{Nucl. Phys. #1}
\newcommand{\pl}[1]{Phys. Lett. #1}

\newcommand{\prl}[1]{Phys. Rev. Lett. #1}

\def\beq{\begin{equation}}
\def\eeq{\end{equation}}
\def\bea{\begin{eqnarray}}
\def\eea{\end{eqnarray}}
\def\bq{\begin{quote}}
\def\eq{\end{quote}}
\def\ben{\begin{enumerate}}
\def\een{\end{enumerate}}

\def\p{{\bf p}}

\def\x{{\bf x}}

\def\etal{{\it et al.}}
\def\ie{{\it i.e.}}
\def\eg{{\it e.g.}}

\def\tan{{\rm tan}}
\def\cot{{\rm cot}}

\def\e{{\rm e}}

\def\vev#1{\langle #1 \rangle}

\def\Dslash{\not{\hbox{\kern-4pt $D$}}}
\def\dslash{\not{\hbox{\kern-2pt $\del$}}}

\def\cp1sub{\setlength{\unitlength}{8pt}\begin{picture}(2,1)\mbox{\scriptsize CP} \end{picture}}                                 

\makeatletter
\@addtoreset{equation}{section}
\makeatother

\title{
\vspace*{-2.0cm}
\begin{flushright}
\normalsize{
FERMILAB-Pub-01/235-T \\
hep-ph/0108199
}
\end{flushright}
\vspace*{1.0cm}
{Neutrinos as the messengers of $CPT$ violation}
\vspace*{0.8cm}
\author{\large\textbf
{G.~Barenboim$^a$, L.~Borissov$^b$, J.~Lykken$^{a,c}$, A.Yu.~Smirnov$^{d}$}\\ 
\\
$^a$\normalsize\emph{Fermi National Accelerator Laboratory,
P.O. Box 500, Batavia, IL 60510, USA }\\
$^b$\normalsize\emph{Columbia University, New York, NY, 10027, USA}\\
$^c$\normalsize\emph{Enrico Fermi Institute, Univ. of Chicago, 5640
S. Ellis Ave., Chicago, IL 60637, USA}\\ 
$^d$ \normalsize\emph{International Center for Theoretical Physics,
34100 Trieste, Italy }\\}
}

\begin{document}
\maketitle

\vspace*{2cm}

\begin{abstract}
$CPT$ violation has the potential to explain all three
existing neutrino anomalies without enlarging the neutrino sector.
$CPT$ violation in the Dirac mass terms of the
three neutrino flavors preserves Lorentz invariance, but generates
independent masses for neutrinos and antineutrinos. 
This specific signature is strongly motivated by braneworld
scenarios with extra dimensions, where neutrinos are the
natural messengers for Standard Model physics of $CPT$ violation
in the bulk.
A simple model of maximal $CPT$ violation
is sufficient to explain the exisiting
neutrino data quite neatly, while making dramatic predictions for
the upcoming KamLAND and MiniBooNE experiments. Furthermore
we obtain a promising new mechanism for
baryogenesis.

\end{abstract}

\thispagestyle{empty}
\newpage

\section{Introduction} 
With the final results by the LSND Collaboration
\cite{LSND} consistently indicating evidence of 
$\bar{\nu}_{\mu}$$-$$\bar{\nu}_e$ oscillations
with a large frequency,
we are faced with the fact that the simplest extensions of the
Standard Model cannot accommodate the observed experimental
anomalies in the neutrino sector. 
With three species of neutrinos only two independent mass differences can be
chosen, and only two of the observed ``anomalies'' (LSND, atmospheric,
and solar) can be explained via oscillations. 

There have traditionally been
two ways out of this predicament: (i) turn a blind eye to the
LSND experiment and keep our fingers crossed that it will be 
contradicted by future experiments such as MiniBooNE,
or (ii) introduce additional neutrino species
(sterile neutrino) and achieve this way the needed third mass
difference. However sterile neutrino scenarios
were recently dealt a blow by the dramatic first results from
the SNO experiment \cite{SNO}. While SNO does not
rule out a non-active neutrino component \cite{SNOtheory},
it certainly makes it harder to reconcile
a sterile neutrino which makes the LSND anomaly possible yet
hides so efficiently in the solar and atmospheric neutrino data.

Another theoretical approach to neutrino anomalies
is to introduce new physics into the neutrino sector,
rather than enlarging it.
For example new flavor-changing neutrino
interactions \cite{FCNI} work well to explain the solar data.
However one should keep in mind that these solutions also
fix one of the mass difference degrees of freedom,
and thus do not directly address our basic predicament. 
Similar statements apply to schemes which postulate
violations of Lorentz invariance or the Equivalence
Principle \cite{ColemanGlashow}.
It should also be stressed that new physics solutions for
the LSND case are even harder to find. In particular
many new physics signals capable of explaining LSND
(e.g. exotic decays of muons) should also have been seen
by the KARMEN experiment.

In this letter we point out that $CPT$ violation has the 
potential to explain all three
existing neutrino anomalies without enlarging the neutrino sector.
$CPT$ violation in the Dirac mass terms of the
three neutrino flavors preserves Lorentz invariance, but generates
independent masses for neutrinos and antineutrinos. This additional
freedom, as we shall see, is sufficient to explain the exisiting
neutrino data quite neatly, while making dramatic predictions for
the upcoming KamLAND and MiniBooNE experiments. Furthermore
we obtain a promising new mechanism for
baryogenesis.

The idea that Lorentz invariant $CPT$ violation could be observable
in the neutrino sector was first suggested by Barger \etal\ \cite{Barger}.
More recently, Murayama and Yanagida suggested that $CPT$ violating
neutrino-antineutrino mass differences could explain a possible discrepancy
between LSND results and neutrinos observed from supernova 1987a
\cite{Murayama}. They also observed that $CPT$ violation  
has the potential to explain all three
existing neutrino anomalies without introducing a sterile neutrino.

\section{$CPT$ violation in the neutrino sector}
Our starting point is the hypothesis that the largest contributions
to neutrino masses are $CPT$ violating Dirac mass terms.
If $CPT$ were conserved, Dirac masses would arise from local
Yukawa type interactions of fields. These interactions would involve
the Standard Model left-handed neutrino
complex Weyl spinor fields $\nu_{iL}(t,\x )$, where $i$$=$$1$, 2, 3
labels the three neutrino species in the mass eigenstate basis.
We would also need the Standard Model
complex Higgs field $\phi(t,\x )$, with
$\vev{\phi}$$=$$v$$=$174 GeV denoting the vacuum expectation
value that breaks electroweak symmetry and gives mass to the
charged fermions. We suppress the $SU(2)_L$ index structure.
In addition,
Dirac mass terms for neutrinos require that we introduce
right-handed $SU(2)_L$ singlet complex Weyl neutrino fields
$N_i(t,\x )$.

Any local field theory interaction that is Lorentz invariant will
automatically conserve $CPT$, so in order to discuss $CPT$ violation
we must go to an operator hamiltonian description in momentum space.
Suppressing flavor indices, we can write standard operator expansions
for the static neutrino fields:
\be
\psi(\x ) = \left({\nu_L(\x )\atop N(\x )}\right)
=  {1\over\sqrt{2}}\int {d^3p\over (2\pi )^3}\; \sum_s
\left( a_{\p}^su^s(\p )\e^{i\p\cdot\x}
+ b_{\p}^{s\dagger}v^s(\p )\e^{-i\p\cdot\x} \right)
\ee
\be
\bar{\psi}(\x ) =  {1\over\sqrt{2}}\int {d^3p\over (2\pi )^3}\; \sum_s
\left( b_{\p}^s\bar{v}^s(\p )\e^{i\p\cdot\x}
+ a_{\p}^{s\dagger}\bar{u}^s(\p )\e^{-i\p\cdot\x} \right)
\ee
Here $u^s(\p )$ and $v^s(\p )$, $s$$=$$1$, 2,  form an
orthogonal on-shell spinor basis,
while $a_{\p}^s$ and $b_{\p}^s$ are anticommuting Fock space
operators:
\be
\{ a_{\p}^r ,a_{\p '}^{s\dagger} \}
= \{ b_{\p}^r ,b_{\p '}^{s\dagger} \}
= {1\over E_{\p}}(2\pi )^3 \delta^{(3)}(\p - \p ')\delta^{rs}
\ee
In the massless limit, $a_{\p}^{1\dagger}$, $a_{\p}^{2\dagger}$
create the neutrino components of $\nu_L$ and $N$, while
$b_{\p}^{1\dagger}$, $b_{\p}^{2\dagger}$ create the antineutrino
components.

The free part of the Hamiltonian is diagonalized by our use
of an orthogonal on-shell spinor basis:
\bea
H_0 &= \int d^3x\; \bar{\psi}(\x )\left[ -i\vec{\gamma}\cdot\vec{\nabla} + m
\right] \psi(\x )  \cr 
~~~ &= \int {d^3p\over (2\pi )^3}\;
(\p^2 + m^2) \sum_s \left[ a_{\p}^{s\dagger}a_{\p}^s + b_{\p}^{s\dagger}
b_{\p}^s \right]
\eea
A $CPT$ transformation interchanges the neutrino Fock operators
$a_{\p}^s$ with the antineutrino Fock operators $b_{\p}^s$.
Thus $CPT$ invariance implies that neutrinos and antineutrinos have the
same mass. Conversely, we can break $CPT$ by introducing independent
mass terms for neutrinos and antineutrinos:
\be
H_0 =
\int {d^3p\over (2\pi )^3}\; \sum_s \left[
(\p^2 + m^2)a_{\p}^{s\dagger}a_{\p}^s +
(\p^2 +\bar{m}^2)b_{\p}^{s\dagger}b_{\p}^s \right]
\label{cptv}
\ee
For $m$$\neq$$\bar{m}$ this hamiltonian violates
$CPT$. It cannot be derived from any local field interaction,
since there is no way to find an orthogonal spinor
basis when the $u^s(\p )$ and $v^s(\p )$ spinors would
have to obey on-shell conditions with different masses.
As a result, this $CPT$ violating but Lorentz invariant
extension of the Standard Model
is nonlocal, \ie , in position space
some neutrino anticommutators will be nonvanishing for spacelike
separations. Although non locality may seem pathological,
the only obvious measurement that detects this
pathology is the measurement of the neutrino and
antineutrino masses through oscillations. 

Restoring the flavor indices, we can parametrize the
observable effects of $CPT$ violation
by three real parameters\footnote{In addition to $\tan\beta_i$,
we will in general need four more parameters (three angles and
a phase) to transform the neutrino mass eigenstate basis into
the antineutrino mass eigenstate basis.}
$\tan\beta_i$,
$i$$=$ 1, 2, 3: 
\be
m = \tan\beta\; \bar{m} \quad .
\ee
For $\tan\beta$$=$$0$, only the antineutrino gets mass, while for
$\cot\beta$$=$$0$ only the neutrino gets mass. We will refer
to either of these two limiting cases as ``maximal'' $CPT$ violation.
For $\tan\beta$$=$$\pm 1$, $CPT$ is restored.

\begin{figure}[ht]
\vspace{1.0cm}
\centering
\epsfig{file=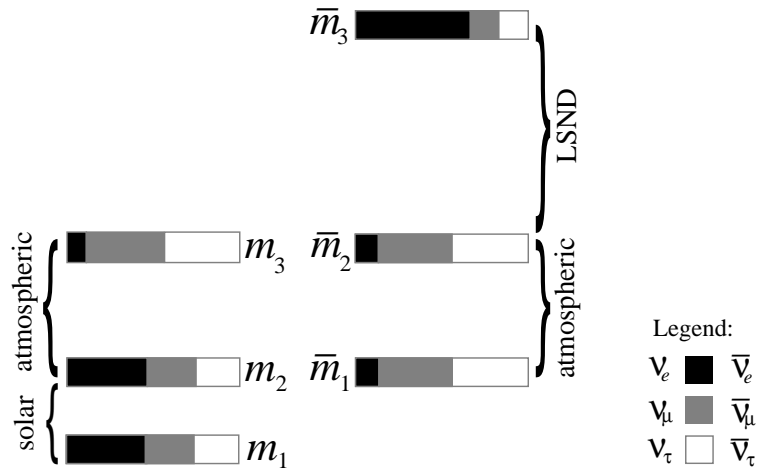,width=10cm}
\caption{\it Possible neutrino mass spectrum in the case of maximal
$CPT$ violation. Although the figure shows an example of large mixing,
our approach is agnostic about the mixing matrix. }
\label{neutrino-spectrum}
\end{figure}

Maximal $CPT$ violation is sufficient to obtain the attractive
neutrino mass spectrum shown in Fig.~\ref{neutrino-spectrum}, which
accounts for the LSND, atmospheric, and solar neutrino data using
only three species of neutrinos. In this simple toy model, 
two antineutrinos, together with one
neutrino, receive $CPT$ violating Dirac masses. The remaining two
neutrinos, as well as the antineutrino, do not receive Dirac
masses, but can generically pick up small masses from higher order
effects, \ie, $CPT$ invariant higher dimension operators like
\be
\nu_L^T \sigma_2 \nu_L \cdot \phi ^* \phi \quad ,
\ee
where again we have suppressed the $SU(2)_L$ index structure.

In the figure, the $\bar{\nu}_{\mu}$ to $\bar{\nu}_e$ transitions
observed by LSND are explained by the large $CPT$ violating 
(dominantly) electron
antineutrino mass. The solar oscillations are the result of the
much smaller $CPT$ conserving mass splittings between $\nu_{2}$
and $\nu_1$. Atmospheric oscillations are assumed to be
$\nu_{\mu}$$-$$\nu_{\tau}$. In the figure,
the $\nu_{2}$$-$$\nu_{3}$
mass-squared splitting has approximately the same magnitude, 
as the $\bar{\nu}_{1}$$-$$\bar{\nu}_{2}$ splitting.
As a result atmospheric muon neutrinos and antineutrinos will have
similar oscillation lengths, in accord with the data from
SuperKamiokande.
Being a water
Cerenkov detector, SuperK does not distinguish neutrinos from
antineutrinos, and washes out any possible difference in
the frequencies of the different channels.

Of course Fig.~\ref{neutrino-spectrum} is just an example:
both spectra are pretty much free and can be accommodated
in many different ways; e.g. one can have inverted spectrum
while the other has normal hierarchy, both can be inverted, etc.
Our approach is agnostic about the mixing matrix.

\section{Mechanisms for $CPT$ violation}

$CPT$ is automatically conserved in a local relativistic quantum field
theory. Possible violations of $CPT$ have traditionally been studied
in tandem with violations of Lorentz invariance \cite{Kost},
with the assumption that
$CPT$ breaking is communicated to all/most sectors of the Standard Model.
In this case the upper bound on the neutral kaon mass difference provides
an extremely stringent bound on $CPT$ violation.

Theoretical motivation for $CPT$ violation usually starts with string theory,
since string theory is not a quantum field theory.
In weakly coupled limits of string theory the low energy effective field
theory will inherit $CPT$ invariance from the $CPT$ symmetry of the
underlying worldsheet dynamics. However it has been suggested that
nonperturbative string effects may violate $CPT$ directly, and it is
also plausible that the choice of string vacuum may violate
$CPT$ spontaneously in the low energy four dimensional effective
field theory \cite{stringycpt}.

We now observe that, in braneworld models of string phenomenology,
the neutrino sector is the most likely messenger of $CPT$ violation
to the rest of the Standard Model. This is because the source of
dynamical or spontaneous $CPT$ violation will lie in the bulk, and
the Standard Model effects will only be visible via couplings of
Standard Model fields (assumed to reside on branes) to suitable
bulk messengers. The generic candidates for the bulk fields which
act as the messengers of $CPT$ are (i) gravity and (ii) the right-handed
neutrinos (\ie , the $SU(2)_L$ singlet neutrinos $N_i$). 
If the extra dimensions are not large enough, gravity effects
are difficult to observe, while the right-handed neutrinos can still have
easily observable Dirac mass couplings. These are precisely the
braneworld scenarios where the effects of Kaluza-Klein sterile neutrinos are
negligible for neutrino oscillation physics. 

Having argued that neutrinos are the likely messengers of $CPT$
violation in a huge class of string models, we may evade all
of the stringent bounds on Standard Model $CPT$ violation from
the kaon sector or any other sector. We may also remain agnostic
on the bulk source of $CPT$ violation, provided that we are convinced
that plausible sources exist.

In this respect it is encouraging to examine a potential simple mechanism.
In realistic braneworld models the Standard Model often resides on a
collection of branes in a higher dimensional spacetime background
which is an orbifold or orientifold. The orbifold background generically
breaks symmetries of the low energy four dimensional effective
description of the Standard Model sector. The broken symmetries can
include spacetime symmetries, \eg , supersymmetry. Scherk-Schwarz type
breaking of symmetries can also occur in such backgrounds.
Thus it is natural to speculate that a suitably contrived 
orbifolding can lead to apparent $CPT$ violation
in the four dimensional effective theory.

More generally, suppose that (by whatever mechanism) some neutral bulk fermion
acquires a $CPT$ violating Dirac mass of the type described in the
previous section. We can then turn on a brane-bulk Yukawa coupling
between the Standard Model $\nu_L$, $\phi$, and
half of the components of this bulk fermion. 
Upon rediagonalization this will communicate the bulk $CPT$ violation
to the observable neutrino spectrum.

\section{Equilibrium baryogenesis}

During the electroweak phase transition in the early universe,
leptons acquire masses from electroweak symmetry breaking. 
A mass difference between neutrinos and antineutrinos would create 
a difference in the chemical potential for populating neutrino and
antineutrino states, resulting in a lepton matter-antimatter
asymmetry proportional to the mass difference. 
This asymmetry is mediated to the baryon sector through
sphaleron processes which violate $B+L$ with great
efficiency.

If we assume that the electron antineutrinos are about 1 eV heavier 
than the neutrinos (as needed to explain the LSND signal)
then the resulting chemical potential between $\nu_1$ and $\bar{\nu}_1$ 
is of the order of 
1~eV. In thermal equilibrium, this will result in a baryon asymmetry given by 
\cite{rocky}
\be
n_B = n_{\nu} - n_{\bar{\nu}} \simeq \frac{\mu_{\nu} T^2 }{6}
\ee
which at the electroweak symmetry breaking scale of 100 GeV gives
\mbox{$\frac{n_B}{s} \sim \frac{\mu_{\nu}}{T} \sim 10^{-11}$}. 
in rough agreement with the observed value.
This mechanism 
does not need CP violation and is produced in equilibrium.

\section{Predictions and discussion}

This $CPT$ violating scenario, with different mass spectra for
neutrinos and antineutrinos, will have dramatic signatures in 
future neutrino oscillation experiments. The most striking 
consequence
will be seen in MiniBooNE (scheduled to
start taking data in 2002),
which is meant to close the discussion about LSND one way
or the other. 
According to our picture, MiniBooNE will be able to confirm LSND
only when running in
the antineutrino mode\footnote{This point was noted already in
Ref.~\cite{Murayama}.}.
Although  their 
original intention was to run primarily in neutrino mode,
the other possibility is under consideration \cite{bill}.
In addition, within this scheme, oscillations of electron neutrinos
driven by $\Delta m_{\mbox{atm}}^2$ are different in neutrino and 
antineutrino channels. In the latter, these oscillations are
strongly suppressed, whereas in the neutrino channel they
can be at the level of the present upper bound. It is important to
notice that in this case the BUGEY bound is irrelevant and large effects
can be expected.

Before that, the  KamLAND detector \cite{kamland}, 
located inside a mine in Japan 
and sensitive only to electron antineutrinos,
will not see an oscillation signal even if the solar neutrinos have
a LMA oscillation pattern. This signature, as well as the
MiniBooNE one, is independent of whether one has the maximal
$CPT$ violating scenario.

Regarding atmospheric neutrinos, all the experiments aimed to
measure the atmospheric mass differences with high precision 
will find that this magnitude is intimately
related to the channel they are exploring, (possibly) discovering
slightly different values for $CPT$ conjugated channels
and even opposite signs. Current experiments, such as SuperK,
do not distinguish neutrinos from
antineutrinos allowing the atmospheric mass difference to be
not necessarily the same in the neutrino and antineutrino
channels. Predictions in this case are not independent of the realization 
of $CPT$ violation (i.e, maximal or not). 

The observation of a neutrino burst from the next supernova
can also provide a useful tool to constrain separately 
both the neutrino and the antineutrino spectra \cite{Murayama}.

This scenario, having a Dirac electron antineutrino mass of 
${\cal O}(1)$ eV, will be explored in the next generation
of beta-decay endpoint searches
such as the proposed KATRIN experiment, 
featuring a large tritium spectrometer
with sub-eV sensitivity \cite{mainz}.

\section{Conclusions}

The general class of models presented here demonstrate that just three
 neutrino flavors with $CPT$ violation can account for all neutrino 
anomalies with oscillations.
These $CPT$ violating models,
which may arise naturally in string theory and brane world scenarios, 
make very specific benchmark predictions that
will be tested in the near future.
An evidence for violation of the $CPT$ symmetry would undoubtedly point
towards more than three spatial dimensions, and will provide an alternative
to Kaluza-Klein mode searches for testing extra dimensions.

$CPT$ violation, which contrary to CP or T violation, can be also seen
in disappearance experiments , 
puts a serious bias on $CP$ violation measurements
which combine 
results for conjugated channels. In that case $CP$ violation is a subdominant
effect while the main effect is $CPT$ violation. In order to measure genuine 
$CP$ violation, at least two detectors for each channel are needed.

\subsection*{Acknowledgements}
\noindent
We thank J. Beacom for endless useful discussions and J. Conrad,
E. Zimmerman, for comments and suggestions.
JL thanks the Aspen Center for Physics for providing a stimulating
research enviroment.
Research by GB and JL was supported by the U.S.~Department of Energy
Grant DE-AC02-76CHO3000 while that of LB by the Sloan Foundation.

\end{document}